\documentclass[sigconf,nonacm,screen,natbib=false]{acmart}

\usepackage[
    datamodel=acmdatamodel,
    style=acmnumeric,
    backend=biber,
    mincitenames=1,
    maxcitenames=2,
    maxbibnames=999
]{biblatex}
\usepackage{csquotes}
\usepackage{tikz}
\usepackage{booktabs}
\usepackage{multicol}
\usepackage{multirow}
\usepackage{makecell}
\usepackage{colortbl}
\usepackage{threeparttable}
\usepackage{paralist}
\usepackage{enumitem}
\usepackage{xspace}
\usepackage{bookmark}
\usepackage{cleveref}
\usepackage{graphicx}
\usepackage[caption=false]{subfig}
\usepackage{rotating}
\usepackage{longtable}

\newlist{questions}{enumerate}{2}
\setlist[questions,1]{label=RQ\arabic*:,ref=RQ\arabic*}
\setlist[questions,2]{label=(\alph*),ref=\thequestionsi(\alph*)}

\begin{document}

\title[Location-Dependent Security and Privacy Perceptions among Activist Tech Users]{``My Whereabouts, my Location, it’s Directly Linked to my Physical Security'': An Exploratory Qualitative Study of Location-Dependent Security and Privacy Perceptions among Activist Tech Users}

\author{Christian Eichenmüller}
\email{christian.eichenmueller@fau.de}
\orcid{0000-0003-1077-7537}
\affiliation{%
  \institution{Friedrich-Alexander-Universität Erlangen-Nürnberg (FAU)}
  \city{Erlangen}
  \country{Germany}
}

\author{Lisa Kuhn}
\email{lisa.kuhn@fau.de}
\affiliation{%
  \institution{Friedrich-Alexander-Universität Erlangen-Nürnberg (FAU)}
  \city{Erlangen}
  \country{Germany}
}

\author{Zinaida Benenson}
\email{zinaida.benenson@fau.de}
\orcid{0009-0006-7158-0219}
\affiliation{%
  \institution{Friedrich-Alexander-Universität Erlangen-Nürnberg (FAU)}
  \city{Erlangen}
  \country{Germany}
}

\renewcommand{\shortauthors}{Eichenmüller et al.}

\begin{abstract}
Digital-safety research with at-risk users is particularly urgent. At-risk users are more likely to be digitally attacked or targeted by surveillance and could be disproportionately harmed by attacks that facilitate physical assaults. One group of such at-risk users are activists and politically active individuals. For them, as for other at-risk users, the rise of smart environments harbors new risks. Since digitization and datafication are no longer limited to a series of personal devices that can be switched on and off, but increasingly and continuously surround users, granular geolocation poses new safety challenges. Drawing on eight exploratory qualitative interviews of an ongoing research project, this contribution highlights what activists with powerful adversaries think about evermore data traces, including location data, and how they intend to deal with emerging risks. Responses of activists include attempts to control one's immediate technological surroundings and to more carefully manage device-related location data. For some activists, threat modeling has also shaped provider choices based on geopolitical considerations. Since many activists have not enough digital-safety knowledge for effective protection, feelings of insecurity and paranoia are widespread. Channeling the concerns and fears of our interlocutors, we call for more research on how activists can protect themselves against evermore fine-grained location data tracking.
\end{abstract}

\keywords{At-Risk Users, Location Data, Security, Privacy, Ubiquitous Computing}

\maketitle

\section{Changing Digital-Safety Challenges}
\subsection{Motivation and Background}
Digital-safety research now extends beyond the conventional ``average user'' assumption to specifically address the risk factors and needs of so-called ``high-risk'' or ``at-risk users''. According to Warford et al. \cite{warford2022sok}, at-risk users encompass individuals with factors that increase their susceptibility to digital attacks and amplify potential harms. At-risk users might be activists, dissidents, journalists, refugees, sex workers, or women seeking an abortion. The changing \emph{technoscape}, a term introduced by Arjun Appadurai \cite{appadurai2001disjuncture} to describe the man-made technological landscape, makes research with at-risk users particularly urgent. From smart home and smart city to smart mobility, digitization is no longer limited to a few personal devices to be switched on and off, but increasingly and continuously surrounds users \cite{dammann2022geographies}. In this new information environment, incessant information processing goes far beyond the awareness or control of the individuals or populations whose data is being processed. Emerging artificial intelligence applications can act as additional accelerators of information processing and pose entirely new privacy risks \cite{lee2024deepfakes}. As a consequence, users with higher security and privacy needs, whilst aware of opportunities offered by new technologies, also face trade-offs regarding their own interaction with these. Existing research shows that at-risk users are aware of new risks and address the ``continuously produced uncertainty'' in relation to data through ``anticipatory data practices'' \cite{kazansky2021depends}. Through exploratory qualitative interviews with activists and politically active individuals, we inquire how security and privacy concerns and respective data practices play out spatially.

\subsection{Related Work}
A growing body of work in HCI addresses at-risk technology users. Important contributions include investigations into digital security experiences \cite{herbert2023digital, nagaraja2009snooping, scott2016security}, collective information security in movements \cite{albrecht2021collective, boyd2021understanding, daffalla2021defensive, guntrum2024keyboard, reisinger2023unified}, and specific digital-safety needs of journalists \cite{ccalicskan2019digital, di2021we, mcgregor2015investigating, mcgregor2017weakest, tsui2021journalists}, academics \cite{tanczer2020online}, refugees \cite{arunasalam2024understanding, simko2018computer}, quasi-public internet personas \cite{samermit2023millions}, sex workers \cite{mcdonald2021stressful}, women seeking an abortion \cite{mcdonald2023did, mcdonald2024threat}, or survivors of intimate partner violence \cite{bellini2023digital, chatterjee2018spyware, freed2017digital, freed2018stalker, matthews2017stories, slupska2022aiding, slupska2021threat}. Providing a scaffolding to our own exploratory investigation of activists' security and privacy concerns, three papers turned out to be guideposts for our research: Warford et al. provide an overview on the existing at-risk user literature \cite{warford2022sok}; focusing on safety challenges for users and researchers, Bellini et al. examine how to conduct at-risk user research safely \cite{bellini2023sok}; Kazansky highlights future-oriented strategies and ``anticipatory data practices'' of at-risk users \cite{kazansky2021depends}.

Mobilizing Bellini et al.'s definition of digital-safety research as ``research about a person's or a group's state of security, privacy, safety, and autonomy, as it relates to their digital footprint'' \cite{bellini2023sok}, and in recognition of the rise of smart environments, which make data traces and digital footprints increasingly hard to control \cite{lin2024moving, jiang2021location}, we conduct an exploratory investigation into the location dependence of security and privacy perceptions among various activists and politically active individuals. The role of location for the (digital) safety of activists is often only acknowledged implicitly, such as when addressing distinct security and privacy challenges during protests or in domestic contexts \cite{wade2021protest}. Research that deals more directly with technical aspects of ``location privacy'' \cite{beresford2003location} often focuses on a specific device type or technology \cite{jiang2021location}. Hence, to the best of our knowledge, so far no study has investigated activists' perceptions from what might be called a ``geo-HCI'' lens -- that is, from a perspective which thinks about HCI through a space-sensitive heuristic. With evermore geolocatable devices in circulation, we contend risk landscapes and risk assessments for activists are changing. This exploratory research advocates for closing a gap in the at-risk user literature by examining how perceptions and strategies of activists are not just anticipatory (that is, oriented towards future outcomes), but how activists increasingly have to manage location and geospatial data revealing their whereabouts in the present.

\section{Methodological Approach}
This section details our approach to prevent research harms. We discuss why we adopt a grounded theory approach and examine the extent to which our approach aligns with or deviates from existing digital-safety research.

\subsection{Safety Issues and Research Harms}
Advocating for ``a safety-conscious approach'', Bellini et al. see research safety as an important and ``specific area of focus in research ethics'' \cite{bellini2023sok}. In their analysis, risks to participants range from breach of confidentiality by researchers, unauthorized access to research data by adversaries, distress, and re-traumatization during the research process, or deanonymization and/or adversarial feedback after publication. Risks to researchers range from burnout and second hand traumatic stress, to intimidation, liability exposure or surveillance by adversaries tracking at-risk users or populations. To better understand safety issues, we endorse transdisciplinary research that involves at-risk users as guides and stakeholders for ethical research. Participatory threat modeling is an example for this approach \cite{slupska2022they, slupska2021participatory}. The primary focus is ensuring the safety of all involved. Since research harms far outweigh the benefits as soon as security and safety are compromised, we continuously reflect each step, and follow secure research protocols and precautions throughout all research phases. For instance, we use a recording device that is not internet-enabled and transcripts are typed out on an air-gapped device. We also abstain from collecting any demographic information from our interlocutors.  

\subsection{Grounded Theory}
We follow a grounded theory approach \cite{charmaz2017constructivist} for two reasons. Firstly, grounded theory suits at-risk user research since heightened safety concerns require continuous reflection. There is no one-size-fits-all formula for at-risk user research, because ensuring safe research conduct necessitates a step-by-step approach. Secondly, grounded theory aligns with the exploratory nature of our research. Given the identified literature gap concerning location-dependent perceptions and strategies of at-risk users, we choose to conduct semi-structured qualitative interviews to explore experiences and perceptions. Insights are derived from back-and-forth conversations that focus on collaboration and dialog.

\subsection{Qualitative Interviews}

\paragraph{Informed Consent and Interviewee-Researcher Relations}
Recruitment of at-risk users is a careful and lengthy endeavor, with first contact and interview often several weeks or even months apart. We focus on participants in areas where we possess some knowledge or have access to individuals who can act as interlocutors, and we are open with candidates about the nature of our research. To ease participant concerns and prevent potential harm, we refrain here from disclosing detailed recruitment procedures. The positionalities of researchers are important for at-risk user research \cite{linxen2021weird, schlesinger2017intersectional, sadi2021problematising}. On the one hand, the characteristics of researchers play a role for participants’ safety perceptions and will influence the data they are willing to provide. On the other hand, positionality also shapes researcher sensitivity, their understanding of an interviewee's context, and what they consider to be ethical conduct. Interviews have been conducted by the first or second author depending on access and familiarity with interviewee background.

\paragraph{Interview Guide and Codebook}
The interview guide (\Cref{app:guide}) is used to prompt reflection on how interviewees protect their security and privacy in daily interactions with digital technologies. It includes sets of question on IT security, assets/data worth protecting, geolocation risks, spaces of risk and protection, as well as on protective measures. Interviews are in either English or German. Cited German statements are translated into English. The actual order of questions depends on interviewee inputs and is adjusted ad hoc to create conversations that feel natural. Audio recordings are used to produce anonymized transcriptions. The codebook (\Cref{app:codebook}) has been developed in iterative rounds of coding with first and second author each coding the same set of interviews. Intermediate consultations have led to codebook adjustments. We have also leveraged a broader team of IT security researchers in an interpretation workshop, where we shared anonymized interview excerpts to reflect on technical details and identify open questions. This report is based on transcription, coding, and analysis of eight interviews. Interviews range from 45 to 90 minutes, with off the record sections (switching off the recording device) on request.

\paragraph{Interviewees}
As Warford et al. \cite{warford2022sok} show, at-risk user research usually targets distinct populations. We are departing from this pattern for two reasons. Firstly, our research question cuts across activist populations. We desire to learn whether politically active individuals from \emph{different} populations share similar security and privacy concerns. Secondly, we suspect that, given the rise of smart environments, the emerging challenges of geolocation and location-dependent risk assessments are shared across activist user populations. Our goal is to highlight issues voiced across different populations, without neglecting the role of user context \cite{malkin2022contextual}. We interviewed two climate activists, an activist fighting against antisemitism, and a journalist, all from Germany, a climate activist from the US, an anti-poverty activist from India, a pro-democracy activist from Iran, and a pro-democracy activist from East Asia. Interviews took place in person. Interviewees are denoted with abbreviations (e.g., I1 for Interviewee 1), or more abstract descriptions (e.g., one activist). The choice is made based on anonymization criteria to prevent profiling.

\section{Results}
The sequence of sub-sections below roughly corresponds to the typical order of themes as they emerge during interview conversations (see also \Cref{app:guide} for interview guide). In the presentation below, device use and data practice is followed by risk perceptions, specific findings on location-based precautions, and geopolitical considerations. 

\subsection{Device Usage and Data Practice}

\paragraph{Personal Devices and Smart Home}
All interviewees use a range of devices. These include laptops, home PCs, tablets, plus one or more mobile phones, often for clearly separated private, professional, and activist purposes. Most interviewees reject smart home devices. One interviewee mentions having used Google Home in the past, until it ``got too uncomfortable''. Although another interviewee owns a smart fan, they find smart devices bothersome because ``they require more detailed attention'' to manage. One activist is already wary of their iPhone’s capabilities and they ``deliberately don't use smart devices''.

\paragraph{Security and Protection Measures}
All interviewees are familiar with encryption and have referenced it in discussions about secure communication. All utilize E2E encrypted messengers (e.g., Signal or Threema), yet for professional reasons some also engage in communication through unencrypted Telegram channels. Almost all interviewees mention using virtual private networks (VPNs). Many interviewees use Tor for web browsing, also some note trade-offs due to usability issues. I3 emphasizes the use of an encrypted external drive to secure data at rest, while I1 has installed additional firewall software on the router. Approaches to handling Proton accounts diverge. I1 and I7, for instance, leverage it for security measures, while I6 opts to delete correspondences originating from Protonmail accounts to protect against theoretical risks emanating from Proton's anonymity features. In sum, all interviewees are security conscious, albeit responses suggest a variety of different approaches and trade-offs in practice. 

\paragraph{Social Media}
All interviewees make use of social media to some extent, yet all are very conscious of the data traces such use generates. As I5 mentions, ``I use my social media only for work purpose. For activist purpose. I don’t share my personal life there.'' I4 and I7 describe how they separate their ``action device'' for protests from their device for social media activity. One interviewee actively uses Facebook and X, formerly Twitter, and is planning to conduct organized social media monitoring of right-wing extremist groups in the future. Another interviewee navigates a continual tension between personal safety needs and follower expectations, noting, ``people not only want to see something of my work, but also of my person.'' To strike a balance, they occasionally share private information but use a ``time delay'' to reduce the risk of attackers ambushing and causing harm.

\paragraph{Anticipatory Data Practices}
Keeping a veil of anonymity in place when participating in protest actions requires planning and what Kazansky calls ``anticipatory data practices'' \cite{kazansky2021depends}. An interviewed climate activist explains how data practices relate to decisions about roles during civil disobedience campaigns. Arrest probabilities are estimated, and data practices are tailored to respective roles. While some abstain from digital interactions to actively engage in an action, others are observers tasked with live-reporting. Another interviewee confirms the stress factors associated with demonstrations, expressing a ``certain relief'' after deciding to no longer participate in such events. However, they highlight that, in their work context, ``a [social media] post critical of the regime'' already creates danger. Due to anticipatory data practices of others around them, this interviewee was asked to ``unfollow'' a family member on social media to prevent harm to career prospects of the family member's child in the future.

\subsection{Risk Perceptions}
\paragraph{General Security and Privacy Concerns}
As at-risk users, some of our interlocutors are on their guard when it comes to common security and privacy threats. I6 mentions they regularly screen an elderly family member's communication for phishing attempts. Expressing the need for heightened caution, they emphasize that they ``need to be extremely vigilant'' due to evermore sophisticated spam emails. They recount an incident involving a phishing email, purportedly an ``invoice by a company'' linked to a recent vacuum cleaner purchase. The realization that all was not right only occurred when they used mouseover to inspect the file path.

\paragraph{Specific Risks to Security and Privacy}
All interviewees recount situations of increased risk exposure, often stemming from their own choices and activities. These situations include attending political protests or mass actions, reporting on the activities of specific ideological groups, or embarking on travels for private or professional reasons. As Warford et al. point out, at-risk users ``have a documented need to verify that new people they encounter online are safe to associate with or to admit into a private space'' \cite{warford2022sok}. In our interviews, caution regarding intimate encounters and self-imposed restrictions are also prominent. In anticipation of future attacks, two interviewees describe curtailing and avoiding relationships. One of them explains they are ``extremely careful'' about getting romantically involved, out of concern for their own security, and for unintentionally exposing others to surveillance targeted at them. 

\paragraph{Heightened Safety Concerns}
Of the eight people interviewed, one stands out as they report receiving several recent anonymous warnings intended to deter their actions. These warnings and their circumstances point towards a powerful adversary with significant capabilities. The interviewee faces tangible personal safety risks, brought to life by numerous anecdotes shared both on and off the record during the interview. That they are not alone with heightened safety concerns comes through in statements by another interviewee. Asked about what kind of data is worth protecting, they immediately respond ``my whereabouts, my location, because it’s directly linked to my physical security''. 

\subsection{Location-Related Precautions}

\paragraph{Controlling immediate surroundings}
Interviewees think about and generally know where devices are located. One interviewee explained how, during meetings involving other at-risk users, the location of devices is often explicitly addressed. One interviewee continuously considers the position of electronic devices in their flat to create non-electronic safe spaces. Another interviewee routinely observes their outdoor surroundings for potential threats, even scrutinizing the shapes of bags and items carried by strangers. Yet, the practice of avoiding or banning devices from their environment depends on the ability to control that environment. When asked in which places they feel safe, one interviewee answers that they even feel tense at home, fearing that someone might have sneaked in and ``installed surveillance'', or that adversaries might have positioned a ``camera outside of my house'' being able to ``see what I'm doing in front of my window''. They state having ``these worries all the time''. To ease similar concerns, another interviewee employs devices to scan their home for bugs, such as ``microphones, FM transmitters, cameras, and red-light devices''.

\paragraph{Managing location data}
Several interviewees maintain clear distinctions between their private and action phones, the latter in some cases equipped with non-personalized or foreign SIM cards. They are highly conscious of where and for which purposes they use each device. When participating in civil disobedience, one climate activist carefully considers traceability. This involves, for example, leaving their mobile device at home or in a location not associated with the action. Interviewees are aware that data collected over longer periods of time also increases risks. For example, the thickness of location data can affect the perceived safety of particular spaces, even if they are deemed safe on first visits. I1 mentions taking detours and changing pathways while walking in the city to avoid creating recognizable spatial patterns of life. I6 highlights a practice of deliberately opting for the least crowded routes, essentially an anti-pattern to the crowds, when heading to protests.

\subsection{Geopolitical Considerations}

\paragraph{Location-Based Safety Perceptions Tied to Geopolitical Conflicts}
The case of journalist Jamal Khashoggi, who was murdered in the Saudi consulate in Istanbul in 2018, is a prominent example of the mortal dangers some at-risk users face, but it is also an example of the geopolitical consequences that accompany attacks on their life \cite{milanovic2020murder}. For dissidents or democracy and environmental activists, there are safer and less safe countries, and this fact plays a big role in structuring safety perceptions \cite{marczak2014governments}. In the interviews, safety concerns repeatedly arise with regard autocratically governed contexts and inadequate legal guarantees. Yet, aside from physical travel, interviewees know that sensitive data from one location might travel without their knowledge -- including to third party governmental authorities. As one user mentions, ``Dictators are simply well connected. They exchange information.'' A number of activists had knowledge of Pegasus, the sophisticated spyware employed by autocratic regimes across the world \cite{deibert2023autocrat, marczak2018hide}. 

How past data practice can invite fears in the present is exemplified by an interviewee sharing their concern about social media activity before traveling to a particular country. This interviewee describes deleting their own posts for fear of repercussions upon crossing the border. Which risk mitigation strategies are attainable also differs depending on context and location. For instance, according to I6, omnipresent CCTV in autocratic contexts causes resignation regarding the challenge of evading surveillance. For them, it renders usually employed precautionary measures, such as scouting CCTV locations before choosing a route, useless.

\paragraph{Threat Modeling and Provider Choices}
Aside from trade-offs when traveling or even moving to another country \cite{tran2023security}, there are also trade-offs in users' technology choices. Threat modeling is helpful to understand the set of considerations structuring day-to-day data practices and user choices \cite{usman2025sok}. In our interviews, it also provides first insights into the stark differences in provider perceptions. One interviewee, a pro-democracy activist from Iran, relies on several Google devices and services, as they regard the chances of their adversary gaining access to sensitive data as minimal. The pro-democracy activist from East Asia shares a similar sentiment. Their primary concern is their adversary, not other third parties. As they explain, ``I would rather trust this kind of big company like Google or Apple. I mean yeah, maybe the US Government is also spying on me, but what would be worse? Being spied on by the Chinese or by the Americans.'' The same user also incorporates considerations of manufacturer choice into their device-related threat model, emphatically stating, they ``definitely won't use Huawei'' devices or those from other Chinese brands due to concerns about potential backdoors or state interference. Another interviewee also considers corporate actors when choosing providers, acknowledging that ``[t]hese services are not necessarily neutral. Rather, they are shaped by individuals and diverse interests that influence their operations.'' They cite Telegram as ``an example of a non-neutral provider, which can be your downfall depending on which side you are on.''

\section{Discussion}
In their review of the at-risk user literature, Warford et al. \cite{warford2022sok} find that at-risk populations have less digital-safety knowledge than they need to mitigate the risks they face. Our interviews substantiate this observation. Thus, below we discuss users' feelings of insecurity, before we propose a schema for analyzing data environments along two axes, and conclude with a reflection on our approach.

\subsection{At-Risk Users' Feeling of Insecurity}

\paragraph{Unknown Unknowns}
The lives of at-risk users play out against the backdrop of invisible information flows and potentially secret operations of their adversaries. All respondents indicate that complete security and safety is illusory and unattainable. In terms of our research, it seems important to acknowledge both, the existence of unknown unknowns, and the psychologically taxing effect this has for at-risk users and populations. 

\paragraph{Known Unknowns}
Almost all respondents describe situations and circumstances that they would like to know more about. Data traces are extremely durable and outlast their personal memories \cite{mayer2011delete}, a fact of which the interviewees are vaguely aware. Data can accumulate over extended time periods, providing a sedimented view on users and their location \cite{kim2020location, thakur2021ambient}. While interviewees are conscious of risks associated with pattern of life analyses, the details of what this might reveal about them remain in the dark. A general difficulty for at-risk users is a lack of knowledge and clarity where data traces of their interactions go. For instance, while there is a general awareness that mobile phones potentially reveal one's location \cite{bitsikas2023freaky, li2018location, thompson2022twelve}, the question with whom such information ends up is much harder to answer and a constant cause for concern. Furthermore, separating fact from fiction can become difficult, as exemplified by one interviewee who had no way of knowing whether rumors about the group collaboration tool Riseup Pad being under constant watch by authorities was an unfounded myth or not. Another person points out that ``the cloud of fear'', which is fed by ``conspiracy narratives'' around technological possibilities of corporate and state actors, must be combated with fact-based strategies to ``remove the breeding ground for anti-Semitic narratives.'' Here, the activist hints at superficial, under-complex projections that can easily transform into antisemitic tropes when surveillance capitalism~\cite{zuboff2019age} is framed as orchestrated by a small, mysterious circle of people.

\paragraph{Paranoia}
The interviewees regularly reflect on paranoia -- a term all of them mention of their own accord. An interviewee recounts an ``intense paranoid phase'' following a security incident. During this period, they vividly remember feeling an overwhelming fear for their life when a scooter passed by, anticipating the driver to ``just drive past and shoot''. Reflecting on their adversaries, an activist characterizes their adversaries' objective as literal ``decomposition''. Simultaneously, and at first glance surprisingly, some interviewees deem themselves ``too insignificant''. Whether this is a factual condition or part of a psychological protection reflex is impossible for us to say. Persistent tension and self-reflection become draining, and interviewees express the need to function. As one person declares: ``You can't live with paranoia.'' While another interviewee acknowledges that ``always every second (...) there might be a security breach'', they conclude that their only recourse is to ``just inform myself as best as possible''. One interviewee from India has emphasized to not allow one's mind to be ``colonized'' by constant anxiety, vigilance and need for resilience.

\subsection{An Analytical Schema for Navigating the \emph{Technoscape}}

\paragraph{Data-Rich versus Data-Poor Environments}
As we realized through conversations with these at-risk activists, implicit threat models are regularly modified depending on the assumed or de facto surrounding technologies. Thus, when evaluating interviewee responses, we can analytically distinguish between data-rich and data-poor surroundings. Environments may be data-rich, with sensors and devices continuously generating information that is passed on elsewhere. Or environments may be data-poor, with minimal to no production of information and/or absent connectivity. At the far end of this spectrum, the latter environment may be a self-made or imposed \emph{digital desert}. Self-made digital deserts, e.g., through jamming, deliberately excluding, or rendering digital devices inoperable in a particular space, can be an attractive option for social movements fearing surveillance, repression or persecution. According to one interviewee, creating a device-free meeting space to reduce attack surfaces is already standard practice to establish a sense of security. However, as shown above, such measures rely on the ability to control one's immediate surrounding. Technological control often lies with higher powers and can potentially be used to the detriment of users, as illustrated by internet shutdowns \cite{bhatia2023protests, akbari2020follow}. Often enough, these large-scale, temporarily imposed digital deserts are used to impede communication between protesters and with the outside world, turning neighborhoods, cities and even entire countries into cut-off, data-poor environments.

\paragraph{Private versus Public Spaces}
Whether a space is private or public matters to at-risk users, as it gives them the ability to control their environment and determines the amount of resources available. At home, users have access to protective resources and can take precautions not feasible in public, such as additional router firewalls. Simultaneously, personal spaces also carry specific risks. Security may be compromised by other household members or users may be geolocated through their devices at home. Available precautionary measures change when entering the home of a friend or acquaintance. The home might have smart vacuum cleaners, security cameras, and voice-controlled virtual assistants. Recording and screening for (un)familiar voices, such devices might introduce dangers into an otherwise safe space. Depending on the nature of personal relationships, guaranteeing by-stander privacy through the selection of technical standards can be subject to negotiations between hosts and their guests \cite{cobb2021would, marky2022you, yao2019privacy}. Removing or turning off devices remains an option, though less feasible than in one’s own home. Public, or semi-public spaces often do not afford such choices at all. For instance, users in train stations or airports have no control over WLAN security or the operation of CCTV and facial recognition systems. For at-risk users this can constrain interaction or lead to outright avoidance. As one interviewee explains, this may include putting on the hood of their jacket when in the field of vision of CCTV or taking extensive detours to avoid certain public areas. Personal risk assessments might even dictate means of transportation. In sum, whether a space is public or private does not determine digital (in)security per se, but by offering or withholding options for risk mitigation and control, it changes the risk landscape and in turn influences how at-risk users navigate such spaces.

\subsection{Methodological Reflection: Towards Location-Aware Digital-Safety Research?}

\paragraph{Transdisciplinarity}
Given the extent to which particular at-risk users' security and privacy perceptions are shaped by technical as well as social, psychological and even geopolitical considerations and circumstances, digital-safety research benefits from a decidedly transdisciplinary orientation. Since it can be hard to navigate the ambiguities of user perceptions, including misconceptions and wrong mental models, and the ongoing debates about fast-evolving technological capabilities, the combination of technical and social science expertise holds the potential for more comprehensive perspectives. Just as some organizations disseminate technical knowledge and provide tools for digital self-defense, better knowledge-sharing between academics and at-risk populations can make a real difference.

\paragraph{Limitations}
This is a report from an ongoing research project and data saturation is not reached. Interviews with at-risk users on location-dependent security and privacy perceptions continue. We thank our interviewees for sharing their intimate and deeply personal risk assessments with us. At the same time, there is no guarantee that interviewees feel comfortable enough to share all they would have to say. A limitation of this line of research is that it operates on trust. In our research, we experience first-hand how geopolitical considerations and anticipation of future difficulties limit the research itself. For instance, while in East Asia, the presence of a researcher's and a potential interviewee's phones in the same room prompted a decision to abstain from conducting an interview. This precaution was meant to avoid that adversaries or authorities retroactively determine the potential interviewee's identity using geospatial data.

\paragraph{Future Work}
Debates about geolocation, geospatial and geo-linked data indicate novel pinpointing capabilities. For those confronted by powerful adversaries, knowledge about capabilities and counter-measures can make the difference between life and death, freedom and imprisonment. Thus, we call for more research on how users can protect themselves against evermore fine-grained location data tracking. For instance, future research could more explicitly address location-tailored privacy-enhancing technologies and expand existing research on built-in device security features, such as Apple's Lockdown Mode \cite{mader2024blame}, by focusing on how these allow for the management of location data.

\section{Conclusion}
The urgency of digital-safety research for at-risk users is undeniable. Digitization is no longer limited to a few personal devices to be switched on and off, but increasingly and continuously surrounds users. The lack of visibility and control over information processing has introduced new uncertainties into at-risk users' daily lives. Drawing on eight qualitative interviews, we contend that the rise of smart environments has posed a number of new challenges, including having to deal with location data and geolocation risks. The response of at-risk users includes attempts to control one's immediate technological surroundings and to more carefully manage device-related location data. For some at-risk users, threat modeling has also shaped provider choices based on geopolitical considerations. Channeling the concerns and fears of our interlocutors, this is also a call for more research on how at-risk users such as activists and dissidents can protect themselves against evermore fine-grained location data tracking.

\section*{Acknowledgments}
This work has received support through the Emerging Talents Intiative of Friedrich-Alexander-Universität Erlangen-Nürnberg (FAUeti).

\printbibliography


\newpage

\onecolumn

\appendix


\section{SUPPLEMENT: Interview Guide}\label{app:guide}

\begin{multicols}{2}
\begin{footnotesize}
\subsection{Introduction}
\begin{itemize}
    \item Informed consent
    \item Assure anonymity
    \item Assure careful handling of data - research/scientific purpose only
    \item Research area: ubiquitous computing (internet of things, internet of everything), IT security, information security practices
    \item Research goal: understanding how digitization of everyday environments affects users and their information security
\end{itemize}

\subsection{Opening Questions}
\begin{itemize}
    \item What electronic or digital devices do you carry with you right now? (If applicable: Should they stay in the room?) 
    \item How many electronic devices do you own? How many of these are digital devices?
    \item Do you own smart home devices, such as Alexa, Bluetooth/ WLAN-compatible appliances, robot vacuum cleaner, or smart TV? (If not, why not?)
    \item How many cameras or devices with an integrated camera do you have at home?
\end{itemize}

\subsection{IT Security}
\begin{itemize}
    \item What is your understanding of IT security?
    \item Which people or institutions do you think might be interested in your data or communications?
    \item Would you consider your digital communications and digital data to be adequately protected? (If not, what are your concerns? If yes, why is that?)
\end{itemize}

\subsection{Assets and Data Worth Protecting}
\begin{itemize}
    \item What information would you want to protect? What type of data is worth protecting?
    \item What would you consider most precarious? \emph{(go through list)}
    \begin{itemize}
        \item communication data
        \item contact lists or contact data
        \item location data or geodata
        \item behavioral data (e.g., daily routines, consumer behavior, work/life patterns, media usage)
        \item personal data (e.g., first and last name, date of birth)
    \end{itemize}
    \item Are there certain devices you would want to protect?
    \item What applications do you consider worth protecting?
    \item What details about your surroundings are worth protecting?
\end{itemize}

\subsection{Geolocation Risks}
\begin{itemize}
    \item When it comes to third parties (e.g., state actors, individuals), do you think about your location data?
    \item What do you think of devices or applications that allow geolocation tracking?
    \item How do you perceive spaces in which data traces are often left behind? (e.g., public WLAN, airports, national borders)
    \item Do you think about data traces that remain on your device and show your previous location? (e.g., WLAN networks used in the past)
\end{itemize}

\subsection{Spaces of Risk \& Spaces of Protection}
\begin{itemize}
    \item Which locations/spaces do you perceive as risky in terms of data security? (e.g. demonstrations in public spaces, national borders, embassies/consulates)
    \item What makes these spaces risky? (e.g., technology infrastructures)
    \item Which spaces do you perceive as protected? (e.g. your own home, street cafés, internet cafés)
    \item What do spaces where you feel protected have in common? (e.g. control, anonymity)
\end{itemize}

\subsection{Protective Measures}
\begin{itemize}
    \item Do you protect your browser data? If so, how?
    \item Do you adapt your browser usage to the situation at hand? 
    \item Does anonymity play a role in your Internet use?
    \item Do anonymous accounts play a role for you? 
    \item Do you make use of multiple accounts per platform? (e.g., depending on role: job/private/activist, language: English/German?)
    \item Do you use deception measures? (e.g., spoofing, bypassing authentication and identification procedures)
    \item Do you take measures to minimize data traces?
    \item Do you generally not use certain apps and devices? (If yes, why?)
    \item Do you avoid certain providers (e.g. ISPs, email providers, Android/Apple)? (If yes, why?)
    \item Do you keep technical devices out of certain rooms? (If yes, which devices and which rooms, and why?)
    \item Do you take any other protective measures?
\end{itemize}

\subsection{Self-Assessment of IT Security / Digital Literacy}
\begin{itemize}
    \item On a scale of 1 to 10, how important do you consider IT security to be? (1=unimportant; 10=very important)
    \item How have you learnt about IT security so far? (e.g., media, acquaintances, own experiences)
    \item Do you feel the need to deal more intensively with the topic of IT security?
    \item What kind of resources would be helpful? (e.g., tutorials, blogs, workshops, hackathons, books)
\end{itemize}

\subsection{Outro / End of Interview}
\begin{itemize}
    \item Willing to have a second interview at some future point in time?
    \item Current communication channel okay or need to communicate differently?
    \item Thank you / leave-taking
\end{itemize}
\end{footnotesize}
\end{multicols}

\newpage

\onecolumn

\section{SUPPLEMENT: Codebook}\label{app:codebook}

\begin{longtable}{@{}p{4.5cm}@{\hspace{0.5\tabcolsep}}p{10.5cm}@{}}
\toprule
\toprule
\textbf{Code} & \textbf{Description} \\
\midrule
Introduction & Introduction of the interview (background, informed consent, recording, etc.). \\
Devices\_carried & Anything related to the devices carried. \\
Devices\_owned & Anything related to the devices owned. \\
Devices\_used & Anything related to the devices used. \\
Laptop & Mentioning laptops. \\
Tablet & Mentioning tablets. \\
Camera & Mentioning cameras. \\
Mobile\_phone & Mentioning mobile phones. \\
Smart\_home & Mentioning smart home devices. \\
OTP\_device & Mentioning OTP devices. \\
Telecommunication & Mentioning traditional telecommunication (non-internet). \\
Telecommunication\_internet & Mentioning telecommunication (internet-based). \\
WLAN & Mentioning WLAN. \\
Connecting\_to\_the\_internet & Mentioning connecting to the internet. \\
E-Mail & Mentioning E-Mail. \\
Social\_media & Mentioning social media (or online engagements beyond writing messages). \\
Provider\_browser\_Firefox & Mentioning Firefox browser. \\
Provider\_browser\_Safari & Mentioning Safari browser. \\
Provider\_browser\_Tor & Mentioning Tor browser. \\
Provider\_e-mail\_Gmail & Mentioning Gmail. \\
Provider\_e-mail\_Hotmail & Mentioning Hotmail. \\
Provider\_e-mail\_Protonmail & Mentioning Protonmail. \\
Provider\_e-mail\_Web.de & Mentioning Web.de. \\
Provider\_messenger\_Signal & Mentioning Signal. \\
Provider\_messenger\_Telegram & Mentioning Telegram. \\
Provider\_messenger\_Threema & Mentioning Threema. \\
Provider\_messenger\_Whatsapp & Mentioning Whatsapp. \\
Assets\_apps & Apps that are considered precious and must be protected. \\
Assets\_data\_at\_rest & Backed up data (hard drives, etc.) that is considered precious and must be protected. \\
Assets\_data\_behavior & Behavioral data that is considered precious and must be protected. \\
Assets\_data\_communication & Communication data that is considered precious and must be protected. \\
Assets\_data\_contacts & Contact data that is considered precious and must be protected. \\
Assets\_data\_location & Location data that is considered precious and must be protected. \\
Assets\_data\_payments & Payment data that is considered precious and must be protected. \\
Assets\_data\_spatial & Spatial information or details describing a place that must be protected. \\
Assets\_devices & Devices that are considered precious and must be protected. \\
Assets\_information & Information that is considered precious and must be protected. \\
Assets\_personal\_information & Personal information that is considered precious and must be protected. \\
Authentication & Mentioning authentication (processes). \\
Encryption & Mentioning encryption (processes). \\
Firewall & Mentioning firewalls. \\
VPN & Mentioning VPN. \\
CCTV & Mentioning CCTV and forms of video surveillance. \\
Facial\_recognition & Mentioning facial recognition. \\
Activism & Mentioning activism. \\
Journalism & Mentioning journalism. \\
Profession & Mentioning profession or professional needs. \\
Private\_life & Mentioning private life. \\
Public\_versus\_private & Mentioning distinctions between public and private life. \\
Attackers & Mentioning attackers. \\
Actors\_on\_the\_scene & Mentioning actors or witnesses. \\
Allies & Mentioning allies or helpful actors. \\
Political\_system & Mentioning political system or political context. \\
Democracy & Mentioning democracy. \\
Autocracy & Mentioning autocracy or autocratic systems and networks. \\
Concentration\_of\_power & Describing concentrations of power. \\
Corporate\_power & Describing corporate power. \\ 
Data\_as\_commodity & Describing data as a commodity. \\
Geopolitics & Mentioning geopolitics or geopolitical constellations. \\
National\_security & Mentioning national security. \\
Surveillance & Mentioning or describing surveillance. \\ 
Attack\_general & Mentioning attacks. \\
Attack\_specific & Mentioning or describing actual attacks. \\
Phishing & Mentioning or describing phishing attacks. \\
Spyware & Mentioning spyware (excluding Pegasus). \\
Spyware\_Pegasus & Mentioning Pegasus spyware. \\
Key\_incident & Mentioning key incidents in attack reconstructions. \\
Attack\_vector & Mentioning or describing attack vectors. \\
Detection & Mentioning or describing detection of malware or malicious devices. \\
Detection\_tests & Mentioning or describing detection tests (e.g. by Citizen Lab or Amnesty Int.). \\ 
TM\_analog & Analog situations or actions that are part of a threat model. \\
TM\_anticipatory\_data\_practice & Strategies and data practices that anticipate potential future (mis)use of data. \\
TM\_apps & Apps that are part of a threat model. \\
TM\_browser & Browsers that are part of a threat model. \\
TM\_data\_continuity & Data that might linger on for a long period of time and is part of a threat model. \\
TM\_device & Devices that are part of a threat model. \\
TM\_OS & Operating systems that are part of a threat model. \\
TM\_risk\_diversification & Behaviors that spread out risks (e.g., using apps/browsers only for certain things). \\
TM\_risk\_strategy & Strategies that deal with risks as part of a threat model. \\
TM\_side\_effects & Side effects in the context of a threat model. \\
TM\_tools\_tech & Technical tools employed in the context of a threat model. \\ 
TM\_tools\_non-tech & Non-technical tools employed in the context of a threat model. \\
TM\_WLAN\_home & Home WLAN that is part of threat model. \\ 
TM\_WLAN\_public & Public WLAN that is part of a threat model. \\
Physical\_safety & Mentioning or describing aspects that contribute or detract from physical safety. \\
Feeling\_of\_safety & Mentioning or describing behaviors or situations that impact safety perceptions. \\
Risk\_perception & Mentioning or describing risks or making risk assessments. \\
Space\_risky & Mentioning or describing spaces that are regarded as risky. \\
Space\_safety & Mentioning or describing spaces that are regarded as safe. \\
Relationship & Role of a relationship or impact on a relationship. \\
Support\_among\_at-risk\_users & Mentioning situations in which at-risk users support each other. \\
Anonymity & Mentioning or describing anonymity. \\
Paranoia & Mentioning paranoia. \\
Censorship & Mentioning or describing censorship (including self-censorship) \\
Trust\_mistrust & Mentioning of trust or mistrust. \\
Known\_unknown & Awareness of aspects that are unknown, unclear or hard to grasp. \\
Potential\_contradiction & Descriptions that seem contradictory. \\
Potential\_falsehood & Descriptions that seem doubtful (and would need to be checked) \\
IT\_security\_definition & Defining IT security. \\
IT\_security\_resources & Mentioning resources for better IT security. \\
IT\_security\_training & Mentioning or describing IT security education and training. \\
Outro & Outro of the interview. \\

\bottomrule
\bottomrule
\end{longtable}

\end{document}